\documentclass[fleqn,10pt]{SelfArx} 

\usepackage[english]{babel} 
\usepackage{epsfig}
\usepackage{lipsum} 

\def\apj{Astrophys. J.}
\def\apjl{Astrophys. J. Lett.}
\def\apjs{Astrophys. J. Suppl.}
\def\mnras{Mon. Not. R. Astron. Soc.}
\def\araa{Annu. Rev. Astron. Astrophys.}
\def\aap{Astron. Astrophys.}
\def\aj{Astron. J.}
\def\nat{Nature}

\def\pasp{PASP}
\def\pasj{PASJ}

\definecolor{color1}{rgb}{0.0,0.0,.0} 
\definecolor{color2}{rgb}{1.0, 1.0, 1.0} 
\definecolor{color3}{rgb}{1.0,1.0,1.0} 
\definecolor{color4}{rgb}{0.0,0.0,0.0} 
\definecolor{colorbox}{rgb}{0.925, 0.956, 0.992}
\definecolor{colorheader}{rgb}{0.33,0.41,0.47} 

\usepackage{hyperref} 
\hypersetup{hidelinks,colorlinks,breaklinks=true,urlcolor=color2,citecolor=color1,linkcolor=color1,bookmarksopen=false,pdftitle={Title},pdfauthor={Author}}

\JournalInfo{Nature Astronomy Perspective} 
\Archive{} 

\PaperTitle{AGN outflows and feedback twenty years on} 

\Authors{C.M. Harrison\textsuperscript{1}$^{\star}$, T. Costa\textsuperscript{2}, C.N. Tadhunter\textsuperscript{3}, A. Fl\"{u}tsch\textsuperscript{4,5}, D. Kakkad\textsuperscript{6}, M. Perna\textsuperscript{7}, G. Vietri\textsuperscript{8,9,1}} 
\affiliation{\textsuperscript{1}\textit{European Southern Observatory, Karl-Schwarzschild-Str. 2, 85748
   Garching b. M{\"u}nchen, Germany}} 
\affiliation{\textsuperscript{2}\textit{Leiden Observatory, Leiden University, PO Box 9513, NL-2300 RA Leiden, the Netherlands}} 
\affiliation{\textsuperscript{3}\textit{Department of Physics and Astronomy, University of Sheffield, Sheffield, S3 7RH, U.K.}} 
\affiliation{\textsuperscript{4}\textit{Cavendish Laboratory, University of Cambridge, 19 J. J. Thomson Avenue, Cambridge CB3 0HE, U.K.}}
\affiliation{\textsuperscript{5}\textit{Kavli Institute for Cosmology, University of Cambridge, Madingley Road, Cambridge CB3 0HA, U.K.}} 
\affiliation{\textsuperscript{6}\textit{European Southern Observatory, Alonso de C\'{o}rdova 3107, Vitacura, Santiago, Chile}} 
\affiliation{\textsuperscript{7}\textit{INAF-Osservatorio Astrofisico di Arcetri, Largo Enrico Fermi 5, 50125 Firenze, Italy}} 
\affiliation{\textsuperscript{8}\textit{INAF-Osservatorio Astronomico di Roma, via Frascati 33, 00078 Monteporzio Catone, Italy}} 
\affiliation{\textsuperscript{9}\textit{Excellence Cluster Universe, Technische Universit\"{a}t M\"{u}nchen, Boltzmannstr. 2, D-85748, Garching, Germany}} 
\affiliation{$^{\star}$ESO Fellow: c.m.harrison@mail.com} 


\title{AGN outflows and feedback twenty years on}

\Abstract{It is the twentieth anniversary of the publication of the seminal papers by Magorrian et~al. and Silk \& Rees which, along with other related work, ignited an explosion of publications connecting active galactic nuclei (AGN)-driven outflows to galaxy evolution. With a surge in observations of AGN outflows, studies are attempting to directly test AGN feedback models using the outflow properties. With a focus on outflows traced by optical and CO emission lines, we discuss significant challenges which greatly complicate this task from both an observational and theoretical perspective. We highlight observational uncertainties involved, and the assumptions required, when deriving kinetic coupling efficiencies (i.e., outflow kinetic power as a fraction of AGN luminosity) from typical observations. Based on recent models we demonstrate that extreme caution should taken when comparing observationally-derived kinetic coupling efficiencies to coupling efficiencies from fiducial feedback models.}

\begin{document}
\fontfamily{lmss}\selectfont
\flushbottom 

\maketitle 


\thispagestyle{empty} 


\noindent During the 1950s and 1960s, it was established that a massive and powerful energy source was required to explain the exceptional luminosities generated by a class of extragalactic objects now known as active galactic nuclei (AGN)\cite{Baade54,Burbidge58,Matthews63}. The energy source was heavily debated\cite{Pacholczyk68,Saslaw74} but the prevailing idea was the accretion of matter onto black holes residing within the nuclei of galaxies\cite{Rees84}, that grow at a rate of $\dot{M}_{\rm BH}$ and have bolometric luminosities of,
\begin{equation}
\label{eq:lagn}
L_{\rm AGN}=\frac{\eta_{\rm r}}{\left(1-\eta_{\rm r}\right)}\dot{M}_{\rm BH}c^{2}.
\end{equation}
The inferred very high mass-to-energy conversion efficiency of $\eta_{\rm r}\approx$\,0.1\cite{Soltan82,Yu02} and high black hole masses (i.e., millions to billions that of the Sun) implies that over the lifetime of a typical black hole, the net energy emitted greatly exceeds the binding energy of their host galaxies\cite{Fabian12}.

It was quickly appreciated that the tremendous amount of energy from AGN could influence galaxy evolution. With effective mechanical (via the jets of charged particles observed in some AGN) or radiative coupling, it became clear that they could heat gas in and around galaxies\cite{Binney95,Ciotti97}. Consequently, AGN became a popular explanation for the ``excess energy'' observed in galaxy clusters\cite{Valageas99}. The constant heating of hot gas around galaxies is sometimes referred to as a ``maintenance mode'' of AGN feedback\cite{Fabian12} and now has convincing evidence due to observed X-ray cavities associated with radio jets and lobes\cite{McNamara12}. 

\begin{figure*}
\begin{minipage}{0.55\textwidth}
\centerline{\epsfig{figure=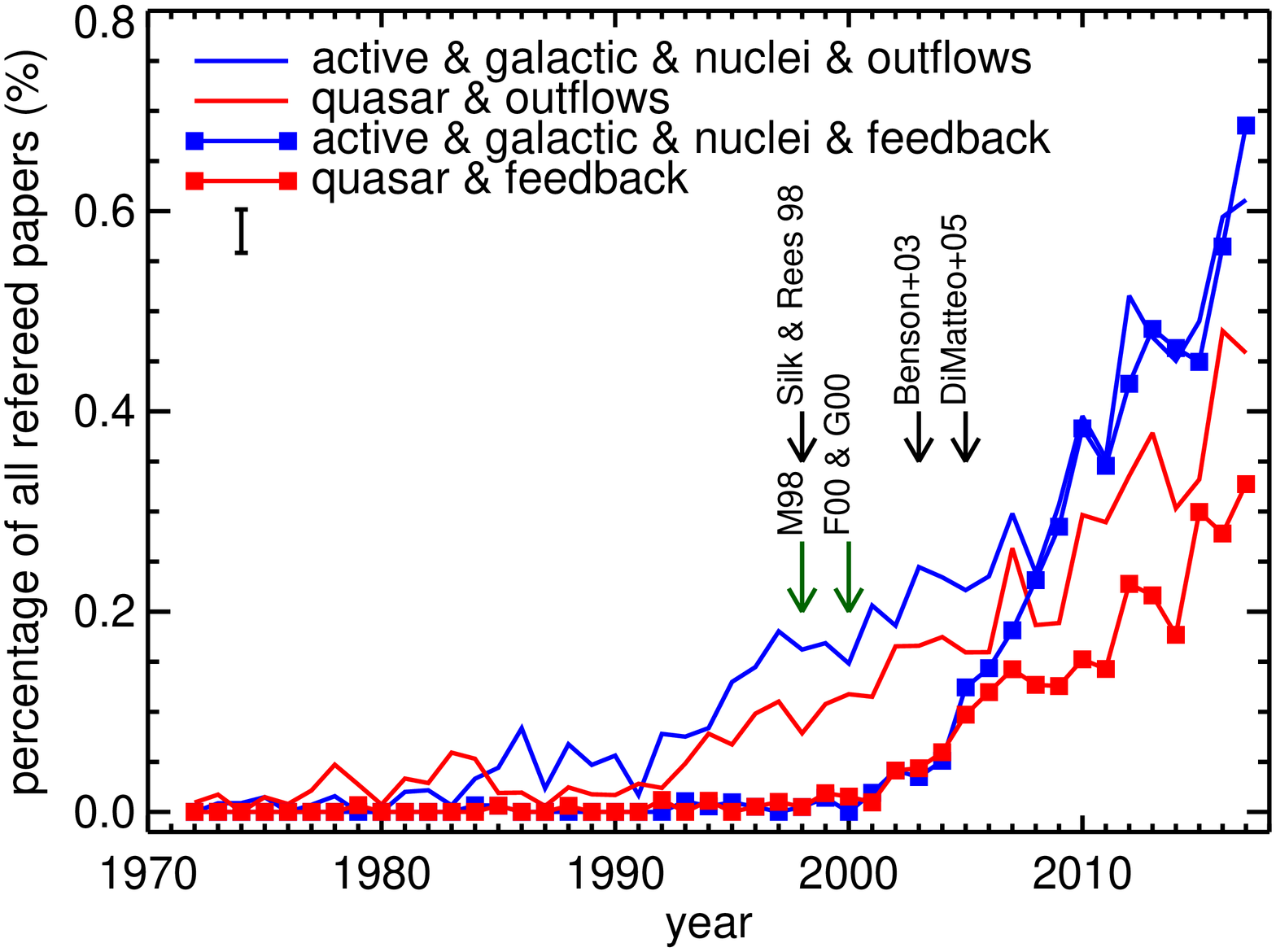,width=0.75\textwidth,angle=90}}\end{minipage}
\hspace{0.2cm}\begin{minipage}{0.42\linewidth}
\fontfamily{phv}\selectfont{\small \textbf{Figure~1 | Abstracts with word combinations of: AGN, quasar, outflows and feedback as a function of time.}

}
\vspace{0.1cm}
{\footnotesize The curves show the percentage of all refereed astronomy publications on SAO/NASA ADS with abstracts containing the combination of keywords shown in the legend, each year. A representative error bar is shown, for $\sqrt N$ errors. The number of abstracts discussing AGN/quasar outflows has been growing relatively steadily since $\sim$1990; however, since $\sim$2000, abstracts mentioning AGN feedback have been growing rapidly. We relate this to: (1) three papers that characterise $M_{\rm BH}$--host galaxy relationships that are in the top-100 of all-time astronomy publications (top 0.009\% citation count based on SAO/ADS) (M98\cite{Magorrian98}; F00\cite{Ferrarese00} and G00\cite{Gebhardt00}) and (2) a series of galaxy formation analytical models (e.g.,\cite{Silk98}), semi-analytical models (e.g.,\cite{Benson03}) and hydrodynamical simulations (e.g.,\cite{DiMatteo05}) where AGN-driven outflows are required to explain these and other observables of galaxy populations. It has become increasingly popular to attempt to observationally derive outflow properties to test AGN feedback models. In this article we discuss the validity of these approaches.

}
\end{minipage}
\end{figure*}

An ``explosion'' of energy from the most luminous AGN (quasars) was also presented from the 1980s as a mechanism to enrich the intergalactic medium and trigger star formation, even to the point of forming entire galaxies\cite{Ikeuchi81,Natarajan97}. Furthermore, in 1988 Sanders et~al. proposed that ultra-luminous infrared galaxies host dust enshrouded quasars whose radiation pressure will subsequently drive high velocity ($\approx$100\,km\,s$^{-1}$), high mass outflows that potentially remove or destroy molecular gas\cite{Sanders88}. In this article we focus on warm ionised and cold molecular outflows, driven by strong AGN radiation fields or radio jets, and their impact upon the host galaxies. This is sometimes referred to as a ``quasar mode'' of AGN feedback\cite{Fabian12}.

It was the period 1998--2000 that AGN feedback, and the role of AGN-driven outflows in galaxy evolution, gained real momentum in the literature (Figure~1). This was initiated by a series of three seminal papers that confirmed earlier suggestions that the masses of black holes ($M_{\rm BH}$) are tightly correlated with the luminosities, velocity dispersions ($\sigma$) and stellar masses of their host galaxy bulges\cite{Magorrian98,Gebhardt00,Ferrarese00}. Starting with Silk \& Rees in 1998\cite{Silk98} (also see \cite{Haehnelt98}) analytical models used simple arguments to show that if only a few percent of a quasar's luminosity is harnessed to drive a wind, they could create a galaxy-wide outflow and establish the scaling relations between the black hole mass and the host bulge properties\cite{Silk98,King03}. Twenty years on, the role of AGN feedback in setting these scaling relations remains fiercely debated\cite{Peng07,AnglesAlcazar17}.

AGN have now became a fundamental component of semi-analytic models and hydrodynamic simulations of galaxy evolution in order to explain the properties of galaxies hosted by massive dark matter haloes ($\gtrsim 10^{12} \, \rm M_\odot$; see review in \cite{Cattaneo09}). In the absence of AGN feedback, models predict galaxies which are too bright, too compact, too massive and too blue at $z \, = \, 0$, compared to the observations\cite{Benson03,Granato04}. Through phenomenological prescriptions for AGN feedback, both simulations and semi-analytic models had success in reducing these problems, generating stellar mass functions, gas fractions and passive galaxy fractions in much better agreement with observations\cite{Benson03,Springel05ellip}.

In order to infer the role of outflows in AGN feedback, observers increasingly attempt to measure properties such as mass outflow rates, $\dot{M}_{\rm out}$, kinetic powers,
\begin{equation}
\label{eq:edot}
\dot{E}_{\rm kin}=\frac{\dot{M}_{\rm out}}{2}\left(v_{\rm out}^{2}+A\sigma_{\rm out}^{2}\right),
\end{equation}
and momentum rates,
\begin{equation}
\label{eq:pdot}
\dot{P}=\dot{M}_{\rm out}v_{\rm out}.
\end{equation}
Here, $v_{\rm out}$ is the outflow velocity and $A$ is a constant which determines the (uncertain) contribution of the velocity dispersion within the outflow, $\sigma_{\rm out}$, to the kinetic power\cite{RodriguezZaurin13}.

In the left panel of Figure~2 we show results from our literature search for observationally-derived outflow kinetic powers presented in the form of ``kinetic coupling efficiencies'', $\dot{E}_{\rm kin}/L_{\rm AGN}$. There is four orders of magnitude variation in the quoted values, that have been derived using a variety of methods, ranging from $\approx$0.001\%  to $\approx$10\% of $L_{\rm AGN}$. Systematic uncertainties in the derived efficiencies are important to understand. These efficiencies are often used to test whether the radiative luminosity of the AGN is capable of driving the observed outflows and to test whether the outflows are as powerful as those suggested by theoretical work to have an impact on galaxy formation\cite{StorchiBergmann10,Holt11,Sanmartim13,Harrison14,Barbosa14,GarciaBurillo14,Feruglio15,Husemann16,Karouzos16b,Bischetti17,Rose17}. Indeed, high values ($\gtrsim$1\%) have sometimes been taken as evidence that AGN outflows are capable of quenching massive galaxies whilst low values ($<$1\%) have been used as evidence that their impact may be limited or that they disagree with theoretical expectations.

\begin{figure*}[!h]
\centerline{\epsfig{figure=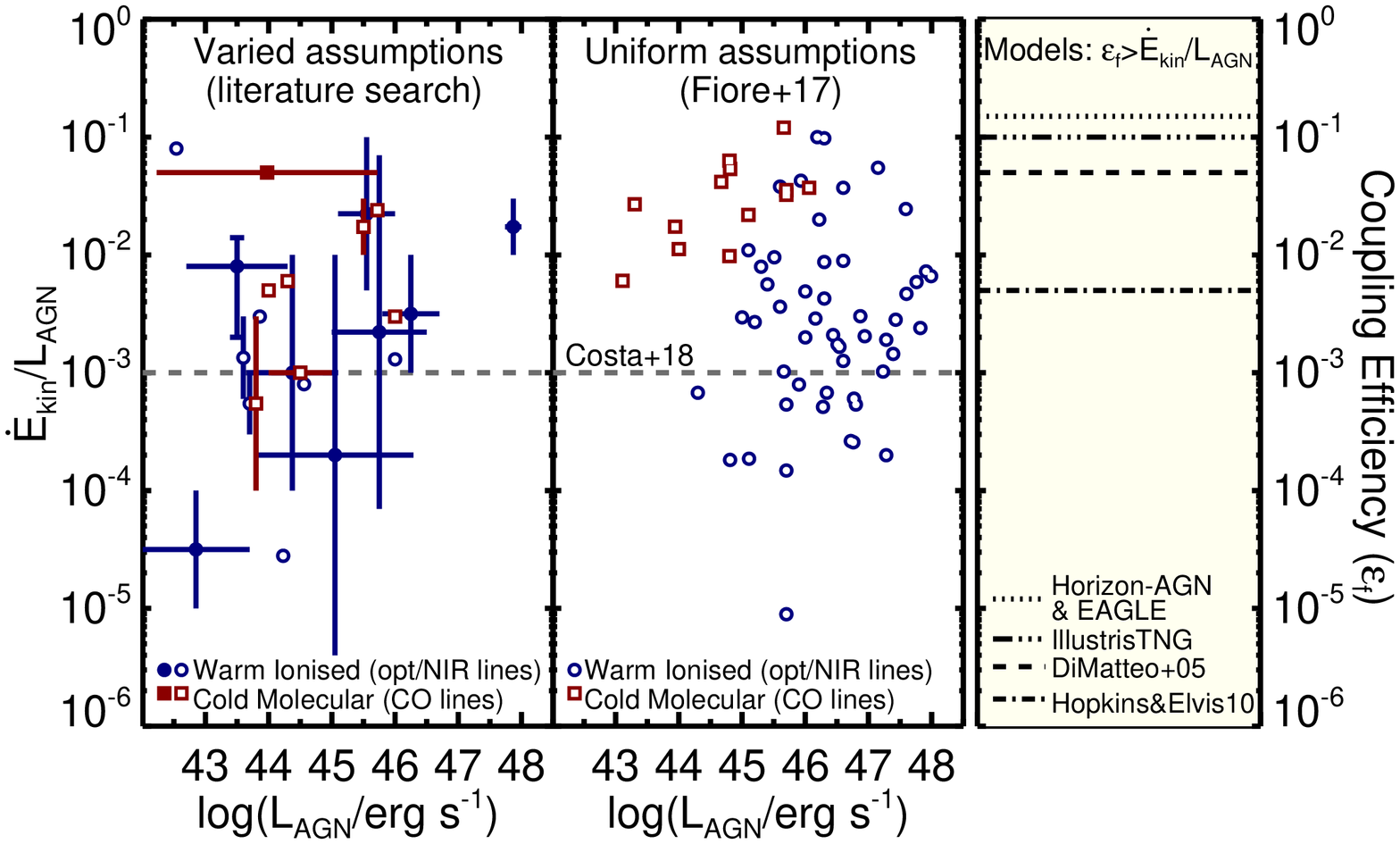,width=0.59\textwidth,angle=90}}
\hspace{0.2cm}\begin{minipage}{\linewidth}
\vspace{0.2cm}\fontfamily{phv}\selectfont{\small \textbf{Figure~2 | Kinetic coupling efficiencies for observed outflows compared to theoretical values.}

}
\vspace{0.05cm}
{\footnotesize A compilation of observationally determined kinetic coupling efficiencies for warm ionised gas from optical or near-infrared (NIR) emission lines (circles) and cold molecular gas from CO emission lines (squares). Hollow symbols are for single galaxies and filled symbols are for galaxy samples. The vertical lines show the range of values quoted, or in the case of an error bar, the quoted error on the average value. The horizontal lines show the range of bolometric luminosity for each sample. In the left panel we show the efficiencies that we found explicitly quoted as the representative/typical values in the abstracts (for warm ionised outflows) or within the papers (for cold molecular outflows) from our ADS search of the ``active \& galactic \& nuclei \& outflows'' subset described in Figure~1 (see Table~1). The middle panel shows the compilation from Fiore et~al.\cite{Fiore17} where kinetic powers were derived from literature observations using a uniform set of assumptions (no errors are provided by the authors). The horizontal dashed line in these two panels shows the ``representative'' (but variable) kinetic coupling efficiency {\em calculated} by Costa et~al.\cite{Costa18}. The right panel shows coupling efficiencies {\em employed} by fiducial AGN feedback models, using the ``quasar mode'' value where applicable\cite{DiMatteo05,Hopkins10,Dubois14,Schaye15,Weinberger17}. In these models, a 100\% of the injected energy is unlikely to become kinetic power in the outflow, so $\dot{E}_{\rm kin}/L_{\rm AGN}<\epsilon_{f}$, and direct comparisons to observed kinetic coupling efficiencies should not be made.

}
\end{minipage}
\end{figure*}

It is not the aim of this article to provide a comprehensive review of the observational and theoretical work on AGN-driven outflows. Instead, focusing on one common method as an example, we highlight some of the observational challenges when deriving kinetic powers. We then assess the validity of the common practise of comparing kinetic powers to ``fiducial'' coupling efficiencies from theoretical models. Throughout we make the assumption that the outflows are truly driven by AGN and not by star formation -  establishing the driving mechanism(s) of outflows is in itself a challenge and is a separate topic of discussion\cite{Morganti18}. 

\section*{Observational results and challenges}
Outflow properties vary as a function of distance\cite{Costa18}; however, only in a small number of objects are outflows sufficiently well resolved, with sufficient signal-to-noise in the emission lines, that the geometries are well known and physical outflow properties can be explored as a function of radius\cite{Crenshaw15,Feruglio15}. In the majority of observations the outflows are marginally resolved, or unresolved, and single average values of outflow properties are usually derived\cite{Liu13,Brusa15,Fiore17}. 

Below we discuss the uncertainties on the average derived outflow properties in these typical cases. Although some of this discussion is relevant for many outflow tracers, we focus on two of the most commonly adopted galaxy-wide (as opposed to accretion disk) tracers, for which large samples over multiple epochs are now becoming available: (1) warm-ionised outflows ($T\approx10^{4}$\,K) traced with rest-frame optical or near-infrared emission lines\cite{Heckman81} and (2) cold molecular outflows ($T\approx 10$\,K) as traced by carbon monoxide (CO) emission lines\cite{Cicone14}.


\subsection*{Mass outflow rates}
Mass outflow rates are usually calculated following
\begin{equation}
\dot{M}_{\rm out} = B\frac{M_{\rm out}v_{\rm out}}{r_{\rm out}},
\label{eq:mdot}
\end{equation}
where $r_{\rm out}$ is the outflow radii and $M_{\rm out}$ is the mass involved in the outflow. The constant $B$ accounts for the assumptions on the geometry and whether an instantaneous or time-averaged mass-outflow rate is being quoted, with $B=$\,1 or $B=$\,3 typically assumed (see discussion in \cite{GonzalezAlfonso17}). In the following, we discuss the ``raw'' outflow quantities of $v_{\rm out}$, $r_{\rm out}$ and $M_{\rm out}$. We focus on Equation~\ref{eq:mdot} as it is a baseline method used by many observational papers, and for studying galaxies from multiple epochs, even when other approaches are also explored\cite{Harrison14,Liu13,Sun17,Fiore17}. Much of the discussion below is also relevant for other approaches that rely on obtaining outflow velocities and radii.

\subsection*{Outflow velocities}
Several approaches are used to define the velocity, $v_{\rm out}$. In an attempt to account for projection effects (also see \cite{Rupke13}), some studies use the velocities of the far wings of the emission-line profiles (e.g. $v_{05}$: the velocity that contains 5\% of the total emission line flux) to infer the true outflow velocity\cite{CanoDiaz12,Rose17}. Alternatively, it is also common to use some combination of the velocity offset of the emission line (with respect to the systemic) and the emission-line velocity width to account for projection effects and infer the de-projected velocities\cite{Liu13,Fiore17}. In most cases these different approaches give similar values of $v_{\rm out}$ within a factor of a few or better\cite{Harrison14,Fiore17}. 

A significant challenge for inferring the velocities at a particular location is beam smearing. In ground-based spatially-resolved spectroscopic data this is dominated by atmospheric seeing. The resulting spatial resolution is typically $\approx$0.5--1.5\,arcsec, corresponding to $\approx$1--3\,kpc at $z=0.1$ or $\approx$4--13\,kpc at $z=2$. However, this can be improved by a factor of a few when using adaptive optics assisted observations and is significantly less of an issue for studying the most nearby AGN\cite{Barbosa09,StorchiBergmann10}. An effect of beam smearing is to artificially broaden emission lines due to the blending of kinematic components. This can have a significant impact on the derived kinetic powers, due to the cubic dependence on velocity (i.e., Equation~\ref{eq:edot} and Equation~\ref{eq:mdot}). Using one particular method on a sample of Type~1 quasars, Husemann et~al.\cite{Husemann16} show that failure to account for beam smearing may lead to kinetic powers that are over-estimated by one-to-two orders of magnitude and find values of $\lesssim$0.1\% of $L_{\rm AGN}$. Another approach for correcting for beam-smearing, applied to spatially-resolved spectroscopic data of some Type~1 quasars, results in inferred kinetic coupling efficiencies that are typically higher and reach up to $\approx$2\% $L_{\rm AGN}$\cite{Rupke17}. The adopted approach significantly affects the conclusions drawn.

\subsection*{Outflow radii}
For warm ionised gas, estimates of $r_{\rm out}$ range from $\sim 10 \, \rm pc$ -- based largely on {\em Hubble Space Telescope} imaging and spectro-astrometry measurements\cite{Rose17} up to $\gtrsim$10 kpc based on ground based spectroscopy\cite{Liu13,Harrison14}. For CO identified outflows, sizes of a few kiloparsecs have also been quoted\cite{Cicone14}. Whilst there are plenty of observations utilising spectroscopy or interferometry that find that AGN-driven outflows are spatially resolved, there is an ongoing debate about whether the outflows are typically sub-bulge scale (i.e, $\lesssim$1--3\,kpc) or more extended\cite{Liu13,Harrison14,VillarMartin16,Husemann16,Rose17,Rupke17}.

The varied conclusions on outflow radii are partly due to different approaches for accounting for beam-smearing (see above). However, some of the differences may also depend on the sample selection (e.g. $L_{\rm AGN}$) and how the outflow sizes are actually determined. For example, outflow sizes can be defined as the spatial extent of ``broad'' ($\sigma_{\rm line}\gtrsim$250\,km\,s$^{-1}$) emission lines or, alternatively the edges of outflowing bubbles \cite{Heckman90,Greene12,Harrison15}. Furthermore, because optical emission lines are illuminated by the radiation of the central AGN, with a very steeply declining radial dependence, it is very difficult to detect the outer regions of warm ionised outflows. In the most challenging cases $r_{\rm out}$ is an upper limit or has to be assumed due to the outflow being completely unresolved in the data\cite{Bischetti17,Rose17}.

\subsection*{Outflow masses}
Determining the mass in a particular gas phase is very challenging. For warm emission-line outflows a common approach is to determine the gas mass from the emission-line luminosity associated with the outflow ($L_{\rm el,out}$) and the electron density ($n_{\rm e}$)\cite{Osterbrock06} following,
\begin{equation}
\label{eq:mass}
M_{\rm out} = C \frac{L_{\rm el,out}}{n_{\rm e}},
\end{equation}
Here $M_{\rm out}$ is the total mass of gas involved in the outflow and in the case of calculations based on the assumption of a mass-conserving shell at a particular radius, this represents the total mass in a spherical or conical volume of radius $r$ with a constant gas density throughout the volume. $C$ is an adopted value that depends on the emission line used and the corresponding assumptions. Ideally hydrogen recombination lines (e.g., H$\beta$ or H$\alpha$) should be used because they are relatively insensitive to the ionisation state and elemental abundances in the outflow. However, $C$ is still temperature dependent\cite{Osterbrock06} and for Type~1 AGN there is the challenge of removing the broad line region contribution to the emission-line profiles. The bright [O~{\sc III}]$\lambda$5007 line is sometimes used as an alternative but the value of $C$ is very uncertain, also depending on the ionisation state and oxygen abundance of the gas\cite{CanoDiaz12}. The value of $L_{\rm el,out}$ can also be uncertain at the factor of a few level, or greater, when the outflows are difficult to isolate in the spectra and/or when the level of extinction is not constrained\cite{Brusa15,Rose17}.

The largest source of uncertainty for warm ionised outflow masses is probably the electron density. Density sensitive emission-line ratios are often used; however, the most commonly used diagnostic ratios of [O~{\sc II}]3726/3729 and [S~{\sc II}]6717/6731 only have diagnostic power in the range $\approx$50--5000\,cm$^{-3}$\cite{Osterbrock06}. Furthermore, due to the small doublet separations the contribution of the outflowing gas to the emission-line profiles can be difficult to isolate, therefore the {\em total} emission-line profiles are often used instead to estimate $n_{e}$. The typical resulting estimates, and/or assumed values, for the outflows of $\approx$\,100--1000\,cm$^{-3}$ (see \cite{Fiore17,Rose17}) may be biased to lower values by the contribution of non-outflowing gas to the emission line ratios\cite{Perna17}. Some studies have instead used transauroral [O~{\sc II}] and [S~{\sc II}] lines, which have a higher critical density. In these cases, higher electron densities are measured in the outflowing line-emitting gas of some nearby AGN ($10^3 < n_{\rm e}  <10^{5.5}$\, cm$^{-3}$), leading to one-to-three orders of magnitude lower mass outflow rates and kinetic powers \cite{Holt11,Rose17}. Similarly high electron densities have been inferred using absorption line outflow diagnostics\cite{Borguet13,Chamberlain15}. However, it is to be verified how sensitive (or not) these measurements are to the underlying elemental abundances and ionization mechanism. Importantly, it is also possible that significant mass and kinetic power is contained in a low density, high-filling factor warm gas component to which all of these line ratios are insensitive\cite{Greene12}. 

To measure the molecular (H$_2$) outflowing gas masses various CO transitions observed at mm and submm wavelengths are often used. The H$_{2}$ mass can be estimated from the CO[1-0] line luminosity ($L_{\rm CO[1-0]}$) using $M(H_2) = \alpha_{\rm CO}L_{\rm CO[1-0]}$\cite{Bolatto13}. As with the ionised outflows, it is difficult to determine what fraction of the line-emitting gas is outflowing; however, the dominant source of uncertainty is probably the choice of conversion factor $\alpha_{\rm CO}$. In most CO outflow studies the $\alpha_{\rm CO}$ value is not measured (although see e.g., \cite{Dasyra16}) and has to be assumed. Typically the optically thick regime is adopted with corresponding values of $\approx$0.8--4, depending on if representative values for local spiral galaxies or mergers are assumed\cite{Bolatto13}. This value is also metallicity dependent and under conditions such as highly turbulent gas motions CO(1--0) may become optically thin, resulting in an order of magnitude lower $\alpha_{\rm CO}$ than is typical for a Milky-Way like disk\cite{Bolatto13,Richings18}. A further challenge, that mostly affects high redshift observations, is that it is common to have to rely on higher excitation CO lines (e.g. CO[2-1], CO[3-2], CO[4-3]) to infer $L_{\rm CO[1-0]}$. The relative luminosities of these lines (i.e., the spectral line energy distribution) often is unconstrained which adds further uncertainty at the factor of a few level.

\subsection*{Discussion of observationally derived kinetic powers}
There is now considerable observational evidence for AGN-driven outflows propagating into the interstellar medium (ISM) of galaxies. Although there is some debate on the best approach to determine the spatial extent and velocity structure of these outflows, the most substantial challenge in obtaining the physical outflow properties, such as the kinetic powers, is probably calculating the gas masses. Overall, for typical observations of barely resolved outflows, the uncertainties on the masses in a particular gas phase and corresponding kinetic powers, can reach a few orders of magnitude\cite{Harrison14,Kakkad16}. Crucially, there is also a deficit of high-quality multi gas phase observations for individual targets, meaning that the {\em total} outflow properties are even more uncertain\cite{Cicone18}.

Some progress has been made to reduce the uncertainties by attempting to directly measure electron densities and $\alpha_{\rm CO}$ values within outflows; however, these measurements remain limited to a few sources\cite{Holt11, Dasyra16,Rose17}. Furthermore, alternative approaches to that described above, have also been used to calculate kinetic powers. One example is to assume a Sedov-Taylor like solution for a supernova remnant where a spherical bubble is expanding into a constant density medium\cite{Nesvadba06,Greene12}. However, the challenges of obtaining the radii and velocities still apply and the assumption of a 100\% filling factor leads to what is likely an upper limit, with 2--3 orders of magnitude higher derived kinetic powers compared to the above method\cite{Greene12,Harrison14,Sun17}. Indirect methods are also now being explored, such as assuming that radio emission traces shocks in the ISM driven by quasar winds\cite{Zakamska14,Harrison15} or by detecting a Sunyaev-Zeldovich signal in the far-infrared spectral energy distributions of quasars\cite{Crichton16}.

The different assumptions and methods used when deriving outflowing gas masses and kinematic luminosities account for at least some of the orders-of-magnitude scatter in the left panel of Figure~2. When comparing different studies (or comparing observations to model predictions), it is important to be aware of any underlying assumptions. Towards mitigating some of these systematic uncertainties Fiore et~al.\cite{Fiore17} compiled literature values of outflow velocities, radii and emission-line luminosities to derive kinetic powers following Equation~\ref{eq:edot} and using the uniform assumptions of: (1) electron densities of $n_{\rm e}=200$\,cm$^{-3}$ for warm ionised outflows; (2) $\alpha_{\rm CO}=0.8$ for molecular outflows; (3) $A=0$ in Equation~\ref{eq:edot}; and (4) $B=3$ in Equation~\ref{eq:mdot}. The resulting outflow kinematic luminosities are shown in the middle panel in Figure~2. Despite uniform assumptions, more than three orders of magnitude range is still observed in the derived kinetic powers for a fixed AGN luminosity for warm ionised outflows. Whilst there is a smaller scatter for molecular values, this is likely to be at least partly driven by an observational bias that currently molecular outflows can only be detected in the most luminous CO emission lines\cite{Cicone18}. A set of uniform assumptions may also be misleading. For example, not accounting for an intrinsic range of electron densities could substantially reduce the accuracy of the quoted values. Furthermore, if the assumed fixed $\alpha_{\rm CO}$ value is unrealistic this would introduce a systematic bias. 

The results of Figure~2 suggest that the varying assumptions in deriving kinetic powers are not completely responsible for the large range of kinetic coupling efficiencies quoted in the literature. Indeed, studies which attempt to accurately measure the densities and radii of the outflows for individual objects -- albeit based on small samples of objects -- find a wide range of coupling efficiencies\cite{Rose17}. Establishing the true intrinsic scatter and how this relates to AGN or host galaxy properties is an ongoing challenge; however, a wide range of measured coupling efficiencies would not necessarily be surprising. For example, it is likely that there is variability in the AGN luminosities and/or the kinetic powers of the outflows\cite{Costa18}. Nonetheless, observational papers often use their inferred kinetic coupling efficiencies to test theoretical expectations, often quoting single fiducial theoretical values\cite{StorchiBergmann10,Holt11,Sanmartim13,Harrison14,Barbosa14,GarciaBurillo14,Feruglio15,Husemann16,Karouzos16b,Bischetti17,Rose17}. Consequently, it is appropriate to now assess the theoretical expectations for kinetic powers from models of AGN feedback.

\section*{Theoretical expectations}
In a first attempt at following AGN feedback in a hydrodynamic simulation of a galaxy merger, di Matteo et al.\cite{DiMatteo05} let black holes, modelled as sink particles, inject thermal energy into their vicinity continuously at a rate $\dot{E}_{\rm feed}$, proportional to their accretion rate, according to
\begin{equation}
\dot{E}_{\rm feed} \, = \, \epsilon_{\rm f} \eta_{\rm r} \dot{M}_{\rm BH} c^2 \, .
\label{eq_dimatteo}
\end{equation}
The product $\epsilon_{\rm f} \eta_{\rm r}$ sets the efficiency at which the accreted rest-mass energy is transferred into surrounding gas resolution elements due to AGN feedback. A term $\left(1-\eta_{\rm r} \right)^{-1}$ in Equation~\ref{eq_dimatteo} is neglected in this and other studies; however, for $\eta_{\rm r} = 0.1$, ignoring this term has a negligible effect (see Equation~\ref{eq:lagn}). The `feedback efficiency' $\epsilon_{\rm f}$, a free parameter, determines the fraction of the AGN bolometric luminosity that couples to gas in the vicinity of the black hole. Di Matteo et al.\cite{DiMatteo05} calibrated $\epsilon_{\rm f}$ to the value required to reproduce the normalisation of the local $M_{\rm BH} \, \-- \, \sigma$ relation. This gave $\epsilon_{\rm f} \, = \, 5\%$, a feedback efficiency which has since become established as a reference value in both theoretical and observational studies of AGN feedback.

Since the scales relevant to black hole accretion and the ejection of AGN winds are not resolved in galaxy formation simulations, how `subgrid' AGN feedback models relate to a fundamental physical driving mechanism is often unclear. 
However, in some instances, fundamental theoretical models do motivate feedback efficiencies of $\epsilon_{\rm f} \approx 5\%$. For instance, King\cite{King03} considers a fast wind launched from the black hole's accretion zone at a mass rate $\dot{M}_{\rm w}$ and a speed $v_{\rm w}$, i.e. with kinetic power $\dot{E}_{\rm w} \, = \,(1/2) \dot{M}_{\rm w} v_{\rm w}^2$. 
Assuming that the energy contained in this nuclear wind is entirely kinetic and that it is driven through Thomson scattering\cite{King15}, i.e. $\dot{M}_{\rm w} v_{\rm w} \, \approx \, L/c$, it follows that, 
\begin{equation}
\epsilon_{\rm f} \, = \, 0.05 \left( \frac{\eta_{\rm r}}{0.1}\right)^{-1} \left( \frac{v_{\rm w}}{0.1c} \right)^2 \, .
\end{equation}
Efficiencies of $\sim 5\%$ may therefore arise naturally if feedback proceeds via the collision between the ISM and an inner wind with speed $\sim 0.1c$, as is the case for observed ultra-fast outflows and broad-absorption line winds\cite{King03}. However, the strong scaling $\epsilon_{\rm f} \,\propto\, v_{\rm w}^2$, implies that a slower inner wind, possibly driven at the dusty torus scale with $v_{\rm w} \, = \, 1000 \, \rm km \, s^{-1}$\cite{Roth12}, may only yield efficiencies of $\epsilon_{\rm f} \, \approx \, 0.006 \%$.

\subsection*{Kinetic coupling efficiencies of outflows}
It is crucial to consider how the energy contained in a nuclear wind is communicated to a galaxy's ISM to form a large-scale outflow. As it collides against the ISM, the wind decelerates through a `reverse shock', while driving a `secondary' outflow consisting of shocked, swept-up, ISM material - the `large-scale outflow'. The large-scale outflow is said to be `energy-driven' if the material composed of shocked wind gas, which pushes onto the shocked ISM component from within, retains its thermal energy, or `momentum-driven' if it radiates it away. Even if the large-scale outflow is energy-driven, only a portion of the original nuclear wind power ends up in kinetic form. Some fraction may be used up in doing work against the gravitational potential, the ambient pressure and, possibly, ram pressure from infalling gas. Consequently, we should expect the total outflow kinetic powers to be smaller than the total energy injection rate such that $\dot{E}_{\rm kin} < \epsilon_{\rm f} L_{\rm AGN}$ (Figure~2).
Indeed, Veilleux et~al.\cite{Veilleux17}, who detect a small-scale wind and a large-scale molecular outflow in the same system, estimate that $\lesssim 0.1$ of the small-scale wind kinetic power is transferred to the molecular outflow.

Using idealised hydrodynamic simulations following the interaction between a fast nuclear wind with $v_{\rm w} \, = \, 0.1c$ and a homogeneous ambient medium, Richings \& Faucher-Gigu\`ere\cite{Richings18b} find that $\approx 0.7 \-- 0.8$ of the mechanical nuclear wind power is transferred to a mostly energy-driven outflowing shell in thermal form. They also estimate that $(0.1 \-- 0.2)\dot{E}_{\rm w}$ is used up in work done against the gravitational potential, leaving only $\approx 0.1 \dot{E}_{\rm w} \approx 0.5\% L_{\rm bol}$ in kinetic form - an order of magnitude lower than the $5\% L_{\rm AGN}$ carried by the nuclear wind in their simulations. However, it remains possible that models based on spherical outflows, which neglect density inhomogeneities that allow for rapid propagation through paths of least resistance, underestimate the kinetic power\cite{Bourne15,Richings18b}.

The similarities between cosmological simulations, employing a subgrid model for AGN feedback with $\epsilon_{\rm f} \, = \, 0.05$, and the King wind scenario, which may predict $\epsilon_{\rm f} \, = \, 0.05$, may be deceptive.  In cosmological simulations, the strength of AGN-driven outflows are sensitive to how AGN feedback is implemented numerically and on the numerical resolution of the simulation. Using a model similar to di Matteo et al.\cite{DiMatteo05}, Weinberger et al.\cite{Weinberger17} show that the thermal energy that is transferred to gas can be lower than that injected by more than order of magnitude due to unphysical radiative losses alone. Various alternative implementations aimed at overcoming numerical radiative cooling losses in simulations have been developed and there is a large variation in the corresponding adopted coupling efficiencies. For example, Booth \& Schaye\cite{Booth09} store up the injected feedback energy until this is sufficient to heat up the surrounding gas above a minimum heating temperature, usually chosen to be high enough that the cooling times are no longer too short. In their fiducial simulations, they employ a feedback efficiency of $\epsilon_{\rm f} \, = \, 0.15$, three times higher than used be di Matteo et~al. Other implementations have included the deposition of momentum or kinetic, rather than thermal, energy around accreting black holes. Choi et al.\cite{Choi12}, who take such an approach, assume $\epsilon_{\rm f} \, = \, 0.005$, one order of magnitude lower than the canonical value, finding stronger feedback than would be achieved with a thermal model alone. 
Crucially, such approaches also reproduce the scaling relations between black hole mass and host properties (see also \cite{AnglesAlcazar17}). Even when winds are launched kinetically, they can undergo shocks and radiate away a substantial part of their energy. There is, \emph{a priori}, no reason why $\dot{E}_{\rm kin} / L_{\rm AGN}$, as inferred from observed outflows, should match the `feedback efficiencies' of AGN feedback prescriptions adopted in state-of-the-art cosmological simulations (Figure~2). 

\subsection*{Multi-phase outflows}
Another difficulty for interpreting theoretical expectations of kinetic coupling efficiencies is the multi-phase structure of outflows. For analytic models and idealised simulations, predictions for the kinetic power of outflows are often based on the whole outflowing material and not on individual phases. However, observers are typically able to access only a specific outflow component, e.g., the cold phase through CO emission or the warm ionised phase through optical emission lines\cite{Cicone18}. In many cosmological simulations, AGN-driven outflows are entirely hot\cite{Barai18}. Furthermore, the multi-phase nature of the ISM affects the prediction of quenching efficiencies\cite{Wagner12}.  Caution is certainly required when comparing observations to theoretical expectations.

From a theoretical point of view there are many questions around the origin of cold molecular material in outflows. It is mostly thought to be associated with gas which is entrained by a hotter wind\cite{McCourt15}, pushed directly through radiation pressure\cite{Zhang17} or gas that cools within the outflow itself\cite{Richings18}. Particularly promising are attempts at directly modeling emission or absorption lines from outflowing cold gas in simulations\cite{Richings18}. These are challenging, requiring very high $\sim \, \rm pc$ resolution, a treatment of the local radiation field and modeling of complex chemistry and are currently too expensive to perform in a cosmological context.

\subsection*{The impact of AGN outflows}
The link between the strength of individual outflows and galaxy quenching is unclear. On the one hand, this is because in cosmological simulations, outflow kinetic powers have generally not been computed. On the other hand, this is because AGN-driven outflows in cosmological simulations may quench galaxies less directly than na\"ively expected.
In particular, even strong, energy-driven outflows may quench only the nuclear regions of star-forming galaxies and an \emph{immediate} impact on the galaxy as a whole may not occur\cite{Costa14}, even if they are important in the long-term. Recently, Costa et al.\cite{Costa18} compared the efficiency of thermally-driven hot bubbles, as may be achieved by the thermalisation of a fast nuclear outflow, with direct radiation pressure on dust at $(0.1 \-- 1) \, \rm kpc$ scales. While thermally-driven outflows have kinetic powers higher by a factor $\gtrsim 5$, their impact on the host galaxy is delayed with respect to that of radiation pressure by $\gtrsim 10 \, \rm Myr$, greater than the characteristic flow times of outflows. Thermally-driven outflows escape the galaxy through paths of least resistance, with limited impact on the ISM, while they accelerate diffuse halo gas efficiently. 

In summary, quenching by outflows may occur through the starvation of the central galaxy as the halo gas accretion rates are reduced, drawing parallels with the `maintenance mode' of feedback. A similar scenario was outlined by McCarthy et al.\cite{McCarthy11}, where quenching was argued to occur via pre-ejection of baryons from the progenitors of current massive galaxies at high redshift. Overall, the timescale over which outflows act, is arguably significantly longer than that for which the AGN is `on'. The real impact of a powerful outflow may occur a long time after the AGN is observed. Consequently, testing feedback models by observationally constraining the long-term negative and/or positive impact on star formation by AGN-driven outflows is an ongoing challenge, with many apparently contradictory results\cite{Harrison17,Cresci18}.

\section*{Conclusions and future prospects}

Since the 1980s AGN have been postulated to play a significant role in galaxy evolution. However, over the last 20 years there has been an explosion of work attempting to connect AGN-driven outflows with galaxy formation (Figure~1) and test theoretical AGN feedback models using derived outflow properties (kinetic powers, mass outflow rates and momentum rates). We have suggested many observational results are uncertain at greater than the order-of-magnitude level. This is concerning because even a change of a factor of 3 (i.e., the choice of $B$ in Equation~\ref{eq:mdot}) can be sufficiently large to change the interpretation of how outflows are driven\cite{Veilleux17}. However, we have also highlighted some recent work that is attacking this problem with a variety of novel approaches.

Despite it being a common practise, we have also shown that it is not trivial to physically interpret the comparison between ``observed'' values of $\dot{E}_{\rm kin} / L_{\rm AGN}$ and theoretical coupling efficiencies employed by early AGN feedback models (Figure~2). Indeed, if large-scale outflows are multi-phase, the inferred values of $\dot{E}_{\rm kin} \approx 5\% L_{\rm AGN}$ for a {\em single} outflow phase from some observations are actually rather puzzling (Figure~2), since they potentially require 10 times higher coupling efficiencies than envisaged even in the most efficient current AGN feedback models. Recent models are now making it possible for a more direct comparison between observed and simulated outflows, by modeling the outflows properties for a variety of chemical species\cite{Richings18b} or by predicting the multi-wavelength signatures of outflows\cite{Nims15}. Recent models also show that ``low'' kinetic coupling values of $\dot{E}_{\rm kin}/L_{\rm AGN}\approx$10$^{-3}$ (i.e., $\approx$0.1\%$L_{\rm AGN}$) can still result in the outflows having a significant impact upon their host galaxies. However, this impact is likely to be delayed and may well happen long after the AGN, or even the outflows, are actually observed\cite{Harrison17,AnglesAlcazar17,Costa18}.

Moving forward we advocate a continuous dialogue between observers and theorists to ensure that the observation-to-model comparisons are valid. Observers can help by tabulating their ``raw'' observed outflow quantities (velocities, emission-line luminosities) and providing observables as a function of radius after taking into account of beam smearing. Where possible, providing host galaxy properties such as black hole and galaxy masses and gas fractions will also aid the comparison to theoretical expectations. Theorists can help by providing direct outflow observables and predicting outflow properties for {\em individual} gas phases. Progress has already been made on both sides and we remain optimistic that this will be the natural way forward based on sophisticated models and the ever improving spatial resolution and sensitivity of observations.

\section*{Acknowledgements} 
We thank the three referees for their constructive comments. We thank the participants of ``The Reality and Myths of AGN feedback'' meeting for inspiring this article and providing constructive comments and the Lorentz Center staff for its organisation. We thank Alex Richings for useful discussions and Wolfgang Kerzendorf for verifying the trends in Figure~1 are reproduced using the full text in all articles on arXiv following the algorithm in \cite{Kerzendorf17}. We acknowledge support from DFG Cluster of Excellence ‘Origin and Structure of the Universe’.




\begin{table}
\begin{tabular}{@{\hskip 0.3cm}c@{\hskip 0.4cm}c@{\hskip 0.4cm}c@{\hskip 0.3cm}c@{\hskip 0.2cm}}
\hline\hline
$\log(L_{\rm AGN})$ & $\log(\dot{E}_{\rm kin}/L_{\rm AGN})$ & Outflow Tracer &Ref.\\
\hline
42.0 -- 43.7&$-$5 --> $-$4&Opt.&\cite{Barbosa09}\\
43.9&$-$2.5&NIR&\cite{StorchiBergmann10}\\
45.7&$-$1.6&CO&\cite{Feruglio10}\\
45.4&$-$2.9&Opt.&\cite{Holt11} \\
43.7&$-$3.5 --> $-$3.0&Opt.&\cite{Sanmartim13} \\
43.6&$-$3.2 --> $-$2.5&Opt.&\cite{Couto13} \\
44.6&$-$3.0&CO&\cite{GarciaBurillo14} \\
46.0&$-$2.5&CO&\cite{Sun14} \\
42.2 -- 45.7&$-$1.3&CO&\cite{Cicone14}\\
45.1 -- 46.0&$-$2.3 --> $-$1.0&Opt.&\cite{Harrison14}\\
42.5&$-$1.1&NIR&\cite{Schonell14} \\
44.6&$-$3.1&NIR&\cite{Barbosa14} \\
44.9&$-$2.2&CO&\cite{Morganti15} \\
45.7&$-$2.0 --> $-$1.5&CO&\cite{Feruglio15} \\
45.8 -- 46.7&$-$4 --> $-$3&Opt.&\cite{Husemann16} \\
43.7 -- 45.1&$-$4 --> $-$2&Opt.&\cite{Karouzos16b}\\
43.6&$-$4 --> $-$2.5&CO&\cite{Zschaechner16} \\
44.2&$-$4.6 --> $-$4.5&Opt.&\cite{Couto17} \\
42.7 -- 44.5&$-$2.1$^{+0.2}_{-0.6}$&Opt.&\cite{Bae17} \\
45.0 -- 46.4&$-$4.2 --> $-$1.2&Opt.&\cite{Sun17} \\
47.7 -- 48.0&$-$2.0 --> $-$1.5&Opt.&\cite{Bischetti17} \\
43.8 -- 46.3&$-$5.4 --> $-$2.1&Opt. \& NIR&\cite{Rose17}\\
\hline\hline
\end{tabular}
\caption{Values and references for data in left panel of Figure~2. $L_{\rm AGN}$ are in erg\,s$^{-1}$. The tracers are: optical (Opt.), near-infrared (NIR) or CO emission lines.}
\end{table}


\end{document}